\documentclass[prd,preprintnumbers,nofootinbib,tightenlines,superscriptaddress]{revtex4}
\usepackage{graphicx}  
\usepackage{xcolor}
\usepackage{dcolumn}  
\usepackage{bm}        
\usepackage{amssymb}  
\usepackage{amsmath}
\usepackage{verbatim}
\usepackage[utf8]{inputenc}    
\usepackage{dcolumn}

\begin{document}

\title{The Structure of Dissipative Dark Matter Halos}

\author{Ran Huo}
\email[]{huor913@gmail.com}
\affiliation{Department of Physics and Astronomy, University of California, Riverside, California 92521, USA}
\author{Hai-Bo Yu}
\email[]{haiboyu@ucr.edu}
\affiliation{Department of Physics and Astronomy, University of California, Riverside, California 92521, USA}
\author{Yi-Ming Zhong}
\email[]{ymzhong@kicp.uchicago.edu}
\affiliation{Kavli Institute for Cosmological Physics, University of Chicago, Chicago,
Illinois 60637, USA}

\begin{abstract}
Dissipative dark matter self-interactions can affect halo evolution and change its structure. We perform a series of controlled $N$-body simulations to study impacts of the dissipative interactions on halo properties. The interplay between gravitational contraction and collisional dissipation can significantly speed up the onset of gravothermal collapse, resulting in a steep inner density profile. For reasonable choices of model parameters controlling the dissipation, the collapse timescale can be a factor of $10\textup{--}100$ shorter than that predicted in purely elastic self-interacting dark matter. The effect is maximized when energy loss per collision is comparable to characteristic kinetic energy of dark matter particles in the halo. Our simulations provide guidance for testing the dissipative nature of dark matter with astrophysical observations.

\end{abstract}

\maketitle

\section{Introduction}
\label{sec:introduction}
The existence of dark matter in the universe is inferred from its gravitational influence on normal matter. Over the past decades, terrestrial dark matter searches have put strong constraints on its interactions with the standard model particles. This leads to speculation that dark matter may reside in a dark sector that contains its own interactions, but couples to the visible sector very weakly if there is any, see, e.g.,~\cite{Alexander:2016aln}. Astrophysical observations can provide important tests for such a scenario. For example, dark matter self-interactions can change inner structure of dark halos, and thus affect motion of luminous matter in galaxies, see~\cite{Tulin:2017ara} for a review. In fact, many galactic observations, such as diverse galaxy rotation curves of spiral galaxies~\cite{Oman:2015xda,Kamada:2016euw,Creasey:2016jaq,Ren:2018jpt,Kaplinghat:2019dhn,Santos-Santos:2019vrw} and stellar kinematics of satellite galaxies in the Milky Way (MW)~\cite{Vogelsberger:2012ku,Valli:2017ktb,Nishikawa:2019lsc,Kaplinghat:2019svz,Sameie:2019zfo,Kahlhoefer:2019oyt}, may favor this self-interacting dark matter (SIDM) scenario~\cite{Spergel:1999mh,Kaplinghat:2015aga}.

Dark matter self-interactions can transport heat in the halo and thermalize its inner region over cosmological timescales, resulting in a large density core, in contrast to a density cusp predicted in the prevailing cold dark matter model~\cite{Dubinski:1991bm,Navarro:1995iw,Navarro:1996gj}. As a self-gravitating thermal system, a field SIDM halo can undergo gravothermal ``catastrophe" if it evolves long enough~\cite{Balberg:2002ue,Koda:2011yb}. In this phase, the self-interactions continuously transport heat out of the center, while its temperature keeps increasing, resulting in a steep density profile. If the self-interactions are purely elastic with the cross section per mass $\sim1~{\rm cm^2/g}$, favored to explain galactic observations~\cite{Tulin:2017ara}, it takes more than $100~{\rm Gyr}$ for a typical field halo to  deeply enter the collapse phase. On the other hand, dark matter interactions can also be dissipative. For instance, if there are multiple states in the dark sector with a small energy gap~\cite{Mohapatra:2001sx,Finkbeiner:2007kk,ArkaniHamed:2008qn,Batell:2009vb,Kaplan:2009de,Loeb:2010gj,Khlopov:2010ik,Frandsen:2011kt,Cline:2012is,CyrRacine:2012fz,Cline:2013pca,Foot:2014mia,Foot:2014uba,Boddy:2016bbu,Finkbeiner:2014sja,Boddy:2014qxa,Schutz:2014nka,Zhang:2016dck,Blennow:2016gde,Braaten:2018xuw,Alvarez:2019nwt}, their collisional excitation and deexcitation in the halo can lead to energy dissipation~\cite{Fan:2013yva,Rosenberg:2017qia,Buckley:2017ttd,Das:2017fyl,Essig:2018pzq}. Ref.~\cite{Essig:2018pzq} shows that mild dissipative dark matter self-interactions can change halo evolution in a dramatic way and significantly speed up the onset of the collapse phase; see also~\cite{Choquette:2018lvq}. Thus, observations of density cores in many dwarf galaxies with low baryon concentration can be used to constrain model parameters that characterize the dissipative interactions~\cite{Essig:2018pzq}.

The previous studies in~\cite{Essig:2018pzq} are based on a modified version of a semi-analytical fluid model~\cite{Balberg:2002ue,Koda:2011yb}. It captures essential features of dissipative SIDM, but the model makes an idealized assumption that the halo is in a quasi-hydrostatic state over the evolution history. Ref.~\cite{Choquette:2018lvq} uses $N$-body simulations, but considers limited parameter space. In this work, we perform $N$-body simulations of dissipative dark matter self-interactions and study their impacts on halo properties in detail. We design a new deterministic algorithm to model the self-interactions and incorporate it into the~\texttt{Gadget2} code~\cite{Springel:2005mi}. The algorithm is validated in the purely elastic limit by comparing our simulations with results from the fluid model that has been well calibrated. We perform dissipative SIDM simulations for dark halos with masses ranging from dwarf to cluster scales, and explore overall features of the halos in dissipative dark matter. Taking a MW halo as an example, we then present simulations with different combinations of model parameters characterizing the dissipative effect and explore the dependence of the collapse timescale on the parameters.  

The paper is organized as follows: In Sec.~\ref{sec:algorithm}, we discuss a particle physics model for dissipative dark matter and introduce our SIDM algorithm. In Sec.~\ref{sec:setup}, we provide information about the setup of $N$-body simulations and initial halo parameters. In Sec.~\ref{sec:profiles} and Sec.~\ref{sec:evolution}, we show simulation results. We conclude in Sec.~\ref{sec:conclusion}.

\section{Particle Physics Model and SIDM Algorithm}
\label{sec:algorithm}

We consider a simple realization of dissipative dark matter, where two dark matter states $\chi$ and $\chi'$ have a small mass splitting~\cite{Finkbeiner:2007kk,Finkbeiner:2014sja,Schutz:2014nka,Alvarez:2019nwt} and they are coupled with a light force mediator~\cite{Feng:2009hw,Tulin:2013teo}. Suppose $\chi$ is slightly lighter than $\chi'$, it is expected that the light state $\chi$ dominates the abundance when structure formation starts~\cite{Blennow:2016gde,Batell:2009vb}; see~\cite{Vogelsberger:2018bok} for an alternative setup. Two types of dark matter self-interactions can occur in a dark matter halo, i.e., $\chi\chi\rightarrow\chi\chi$ and $\chi\chi\rightarrow\chi'\chi'$. We further assume that $\chi'$ produced in the latter up scattering promptly decays back to $\chi$ through emitting a massless species that escapes from the halo without reabsorption. For the up scattering to happen in the halo, the relative velocity between two incoming $\chi$ particles $v_{\rm rel} \equiv |\vec v_1 -\vec v_2|$ must satisfy the following condition,
\begin{eqnarray}
\frac{1}{4} {m_\chi} v_\text{rel}^2 > 2 (m_{\chi'}-m_\chi) c^2,
\label{energycondition}
\end{eqnarray}
where $m_\chi$ and $m_{\chi'}$ are $\chi$ and $\chi'$ masses, respectively, and $c$ is the speed of light. We introduce a threshold velocity defined as 
\begin{eqnarray}
v_\text{th}\equiv c\sqrt{2(m_{\chi'}-m_{\chi})/m_\chi},
\end{eqnarray}
and rewrite the condition Eq.~\ref{energycondition} as $v_\text{rel}> 2 v_\text{th}$. We denote elastic and dissipative scattering cross sections per mass as $\sigma'/m\equiv\sigma_{\chi\chi\to\chi'\chi'}/m_\chi$ and $\sigma/m\equiv\sigma_{\chi\chi\to\chi\chi}/m_\chi$, respectively, and assume both of them are independent of the relative velocity, $v_{\rm rel}$. This is a good approximation for this work, as we are interested in halo dynamics for a given mass scale that has one characteristic dark matter velocity dispersion.

\begin{figure*}[t]
    \centering
  \includegraphics[scale=0.46]{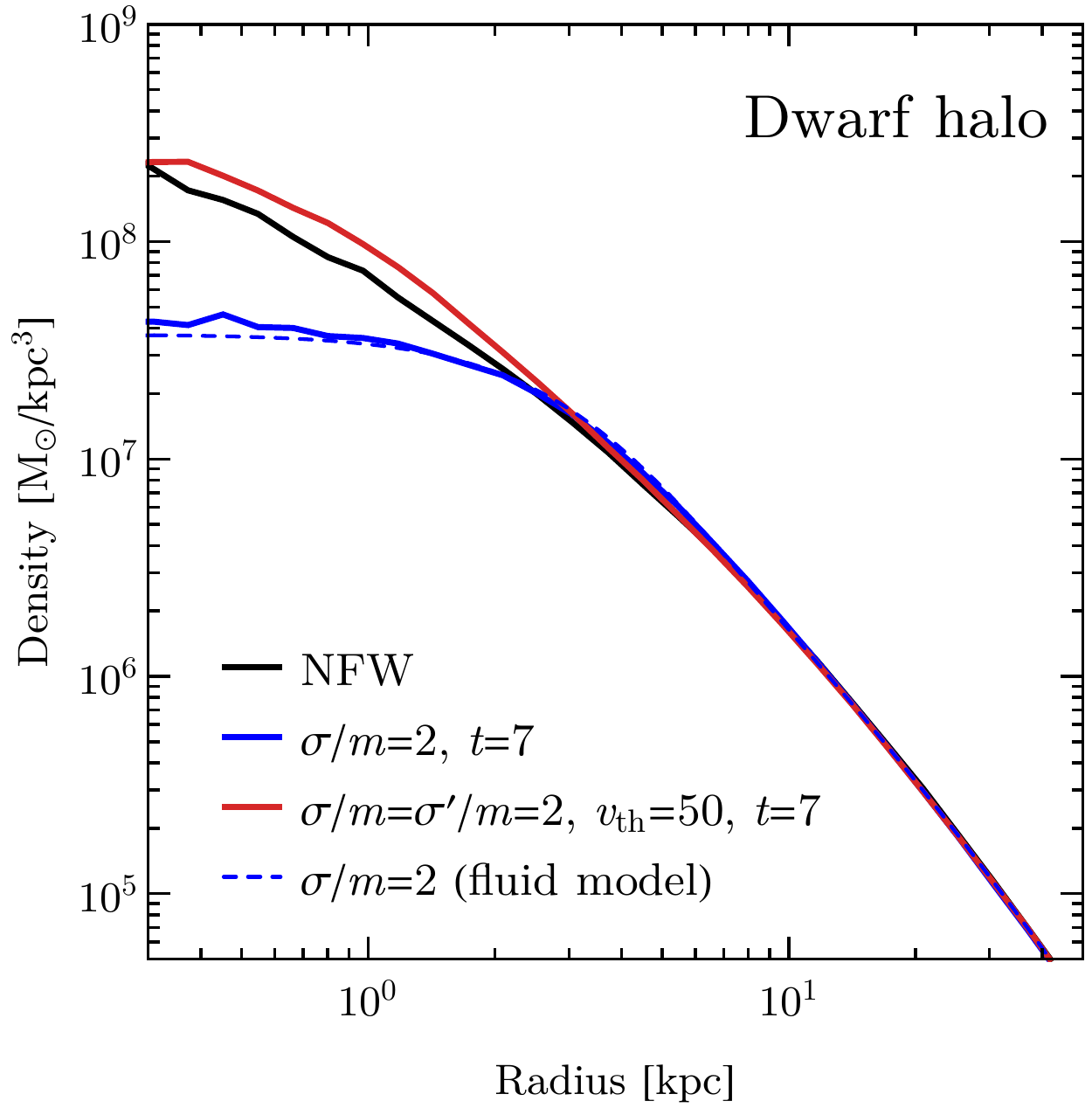}
  \includegraphics[scale=0.46]{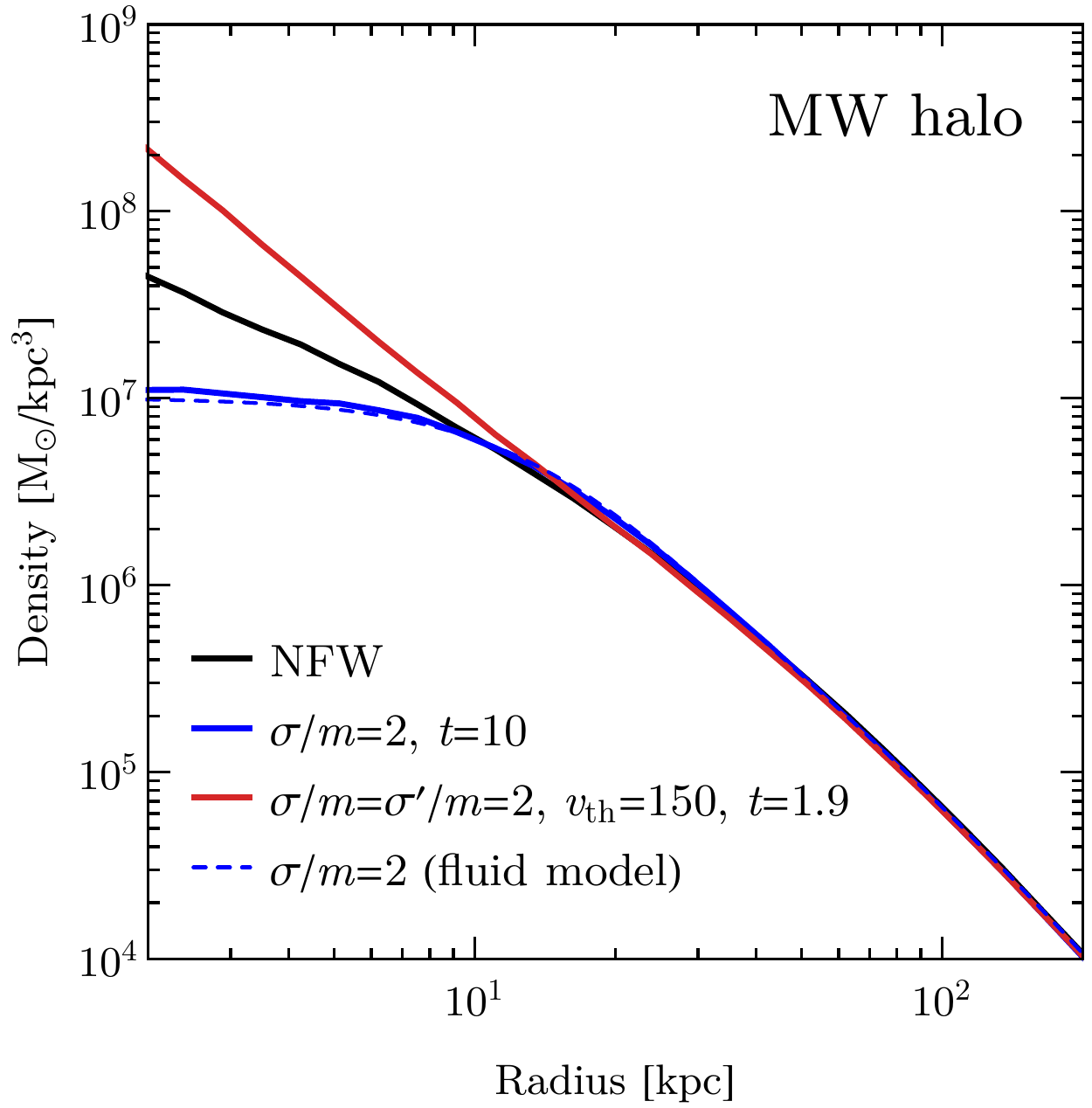}
  \includegraphics[scale=0.46]{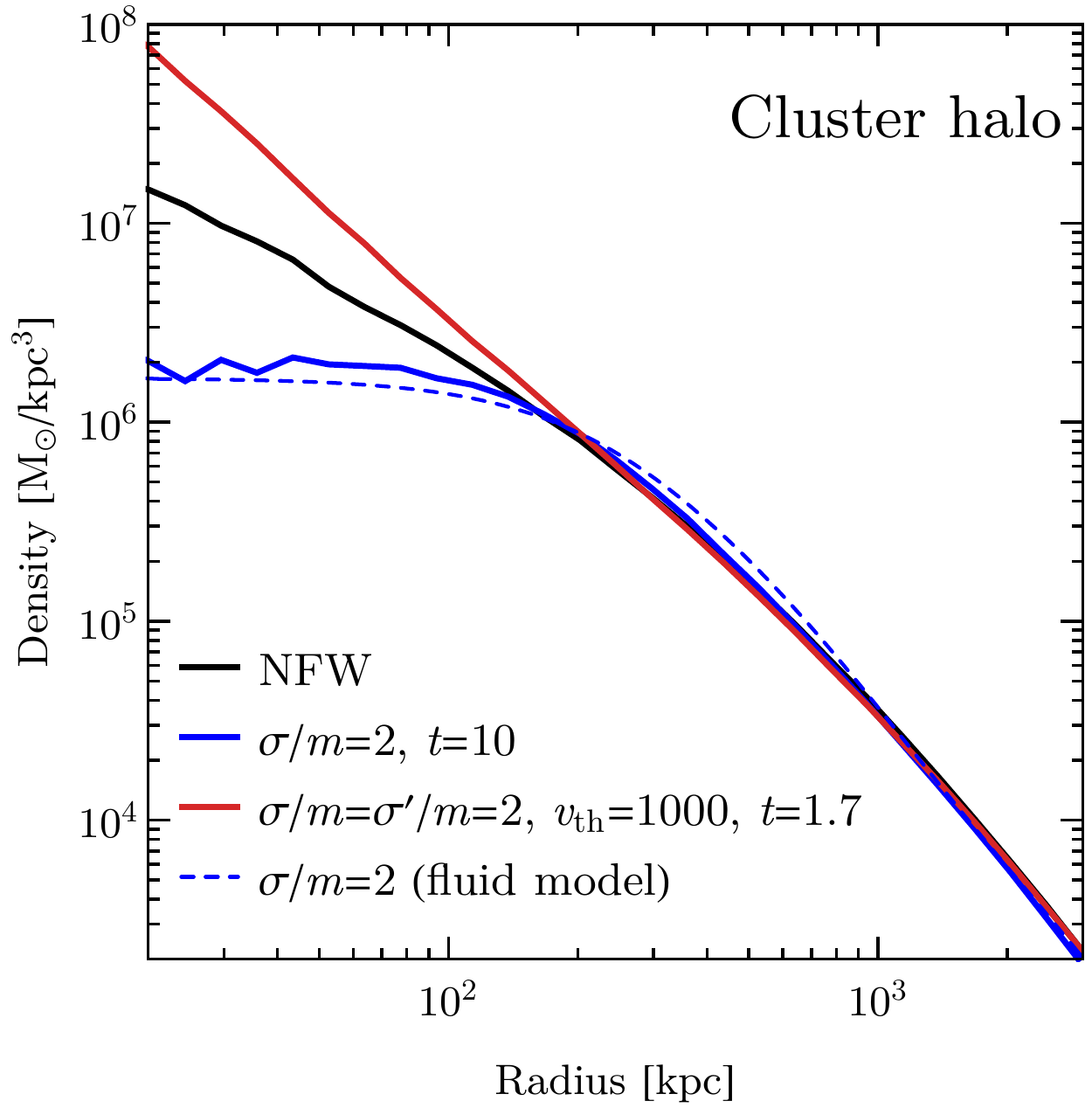}
  \includegraphics[scale=0.46]{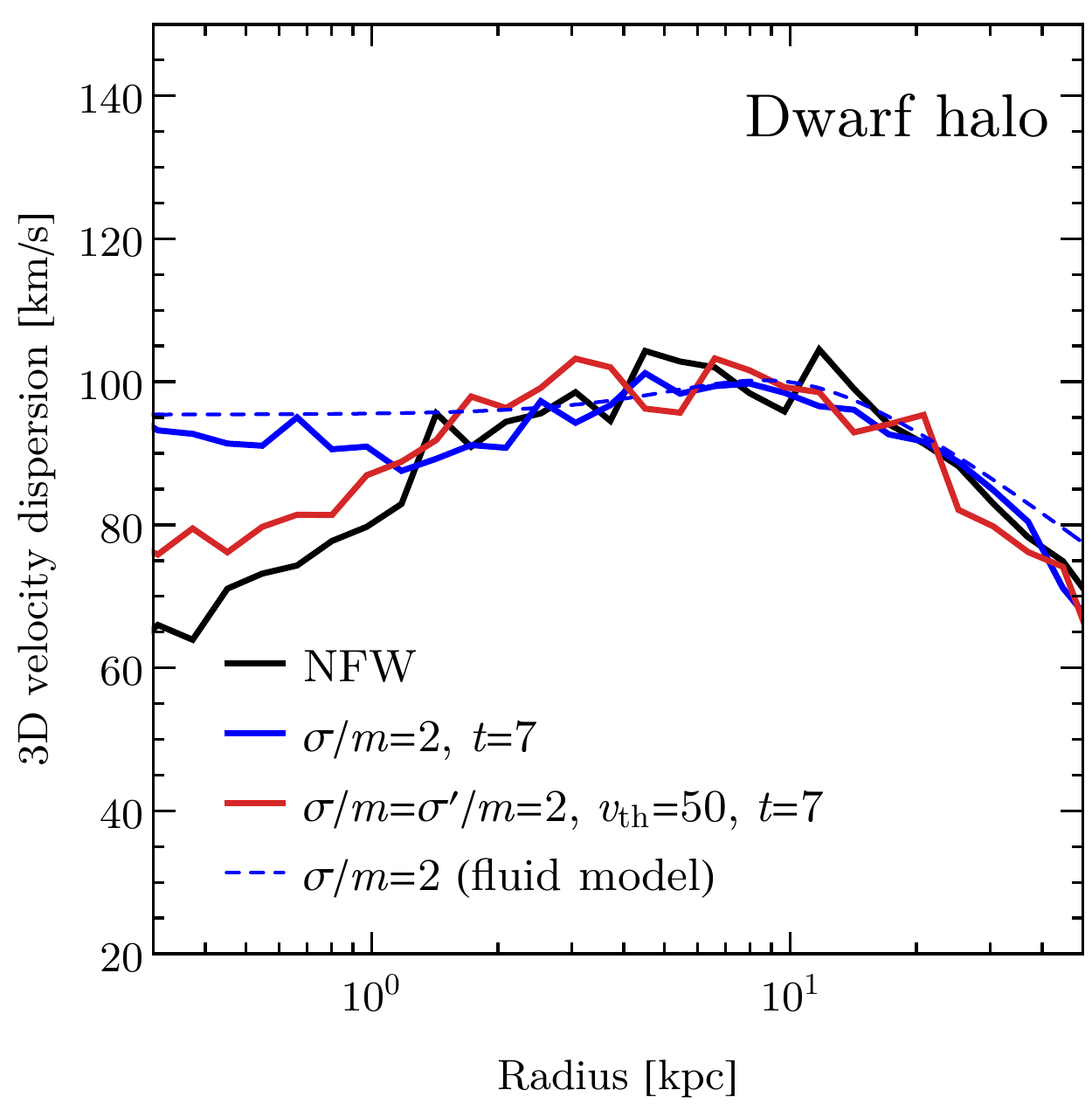}
  \includegraphics[scale=0.46]{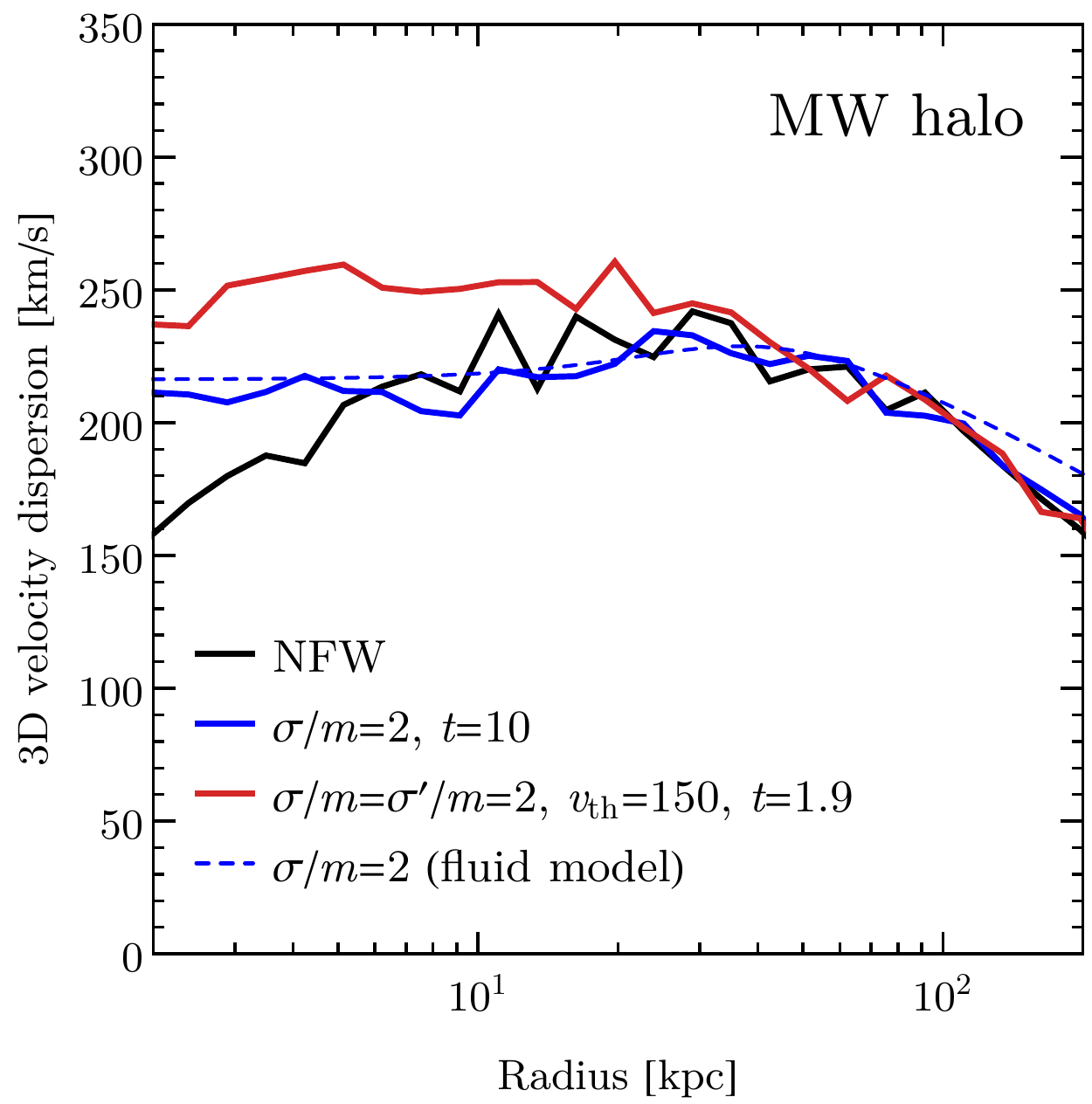}
  \includegraphics[scale=0.46]{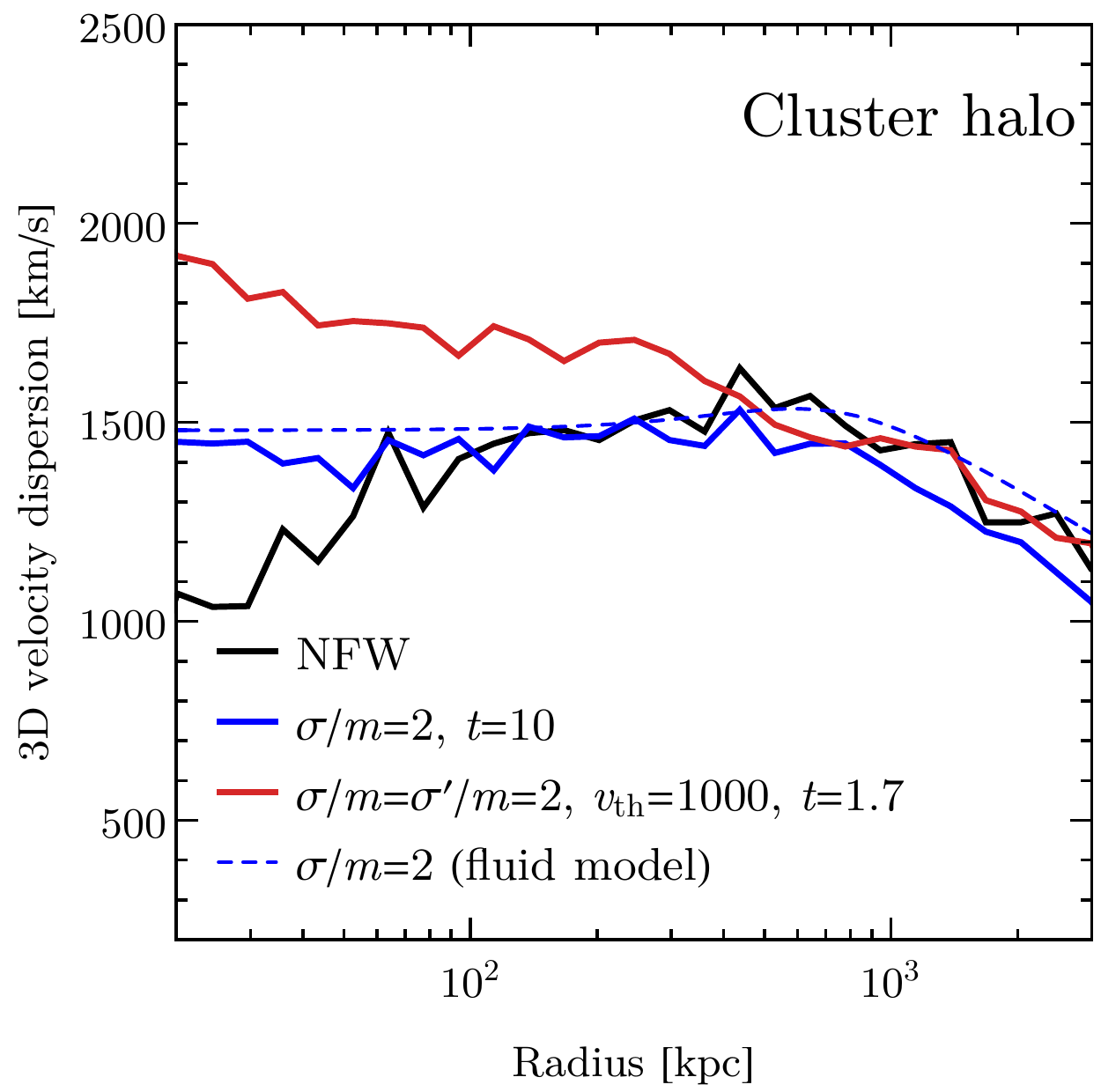}
      \caption{Dark matter density (top panels) and velocity dispersion (bottom panels) profiles for the dwarf (left), MW (middle) and cluster halos (right). We include the initial NFW profiles (black), purely elastic (blue) and dissipative (red) SIDM simulations. If dark matter self-interactions are purely elastic, the inner halo forms a large density core. While, if dark matter has sizable dissipative interactions, the halo can collapse in much less than $10~{\rm Gyr}$, resulting in a steep inner density profile. We also show results from the calibrated fluid model in the purely elastic limit (blue dashed), which agree with our simulations within $10\%$ in density and velocity dispersion. In the legends, the cross section per mass, threshold velocity and evolution time are in units of ${\rm cm^2/g}$, ${\rm km/s}$ and ${\rm Gyr}$, respectively.}
\label{fig:fig1}
\end{figure*}

We design a new algorithm to model dark matter self-interactions in $N$-body simulations and incorporate the associated module into the~\texttt{Gadget2} code~\cite{Springel:2005mi}. We treat each of simulated halo particles as a rigid ball with its ``geometric" radius as
\begin{equation}
\label{eq:ballsize}
r_{\rm p}=\sqrt{\frac{\sigma_{\rm tot}}{m}\frac{M_{\rm p}}{\pi}},
\end{equation}
where $M_{\rm p}$ is the mass of a simulated particle and $\sigma_{\rm tot}$ the total dark matter self-scattering cross section. It is calculated as
\begin{equation}
\sigma_\text{tot}(v_\text{rel}) = \sigma + \sigma' \Theta(v_\text{rel} - 2 v_\text{th}),
\end{equation}
where $\Theta$ is the Heaviside step function. For purely elastic scattering $\chi\chi\rightarrow\chi\chi$, a colliding particle's initial ($\vec{v}_i$) and final ($\vec{v}_f$) velocities in the center of mass frame are the same in magnitude, but differ in directions. While, for upper scattering $\chi\chi\rightarrow\chi'\chi'$, $\chi'$ has a final speed of $v_f = {\sqrt{v_i^2-v_\text{th}^2}}$. Note the up scattering condition $v_\text{rel} > 2 v_\text{th}$ is equivalent to $v_i > v_\text{th}$ in that frame. Our simulations follow the following scattering kinematics
\begin{equation}
  v_f =
    \begin{cases}
      v_i & \text{if $v_i\leq v_\text{th}$}\\
      v_i & \text{with prob. $ = \frac{\sigma}{\sigma+\sigma'}$ if $v_i> v_\text{th}$}\\
      {\sqrt{v_i^2-v_\text{th}^2}} & \text{with prob. $= \frac{\sigma'}{\sigma+\sigma'}$ if $v_i> v_\text{th}$}
    \end{cases}
\end{equation}
In the limit of $\sigma' \to 0$, we recover scattering kinematics for purely elastic self-interactions. We assume the scattering is isotropic in the center of mass frame of two colliding particles.

In our SIDM algorithm, we assume two simulated particles collide once due to dark matter self-interactions if their distance is less than $2r_{\rm p}$ at any given moment. In practice, our module uses a log file to keep track of temporal position and velocity information of all simulated particles between two successive steps of gravity calculations with time interval $\Delta t_{\rm g}$, set in~\texttt{Gadget2}. The flow of execution is as follows: (i) within each $\Delta t_{\rm g}$, the module computes motion trajectories of the particles, and identifies the pair that would collide {\em first in time} based on the collision criterion, i.e., the distance is less than $2r_{\rm p}$, and its collision time $t_1$; (ii) if such a pair exists ($t_1<\Delta t_{\rm g}$), the module assigns individual particles in the pair new velocities with isotropic random directions, subject to constraints of momentum and energy conservations, and updates their information in the log file accordingly; if not ($t_1>\Delta t_{\rm g}$), it directly goes to (iii); repeat this process until the end of the time interval ($\Delta t_{\rm g}$); (iii) the module feeds the gravity solver with the position and velocity information stored in the latest log file, and the solver calculates gravitational interactions. After updating the log file, the module follows the above (i)--(iii) steps when the next gravitational time step starts. 

This algorithm has several interesting features. Since its collision criterion is based on the classical particle trajectories, which are completely determined as we know the initial conditions, our approach is ``deterministic," compared to Monte Carlo-based SIDM algorithms~\cite{Burkert:2000di,Kochanek:2000pi,Yoshida:2000uw,Dave:2000ar,Colin:2002nk,Koda:2011yb,Vogelsberger:2012ku,Rocha:2012jg,Robertson:2016xjh}. In addition, our module is independent of the gravity solver and it handles self-scattering kinematics in time order, a way to evade the multiple scattering problem. Thus, our time step for calculating gravity can be larger than that adopted in the literature. Furthermore, our geometric radius $r_{\rm p}$ defined in Eq.~\ref{eq:ballsize} can be regarded as the ``smoothing length" introduced for calculating the local self-scattering probability in previous algorithms. The difference is that the former is fixed in our algorithm, while the latter needs to be adjusted empirically or dynamically in those based on the Monte Carlo approach. 

To check our algorithm, we have developed an independent code based on the Monte Carlo approach. In the purely elastic limit, where a density core form, both codes produce almost identical velocity dispersion and density profiles for a testing halo. For dissipative SIDM simulations, when the collapse occurs and the self-scattering rate becomes high, it becomes inconvenient to optimize the smoothing length for our Monte Carlo-based code. But the deterministic one does not have this issue as the geometric radius is fixed. We have also checked our new algorithm using the fluid model, which has been carefully calibrated to cosmological SIDM simulations~\cite{Essig:2018pzq}, as we will show later. 

\section{simulation setup}
\label{sec:setup}

\begin{figure*}[t]
  \includegraphics[scale=0.46]{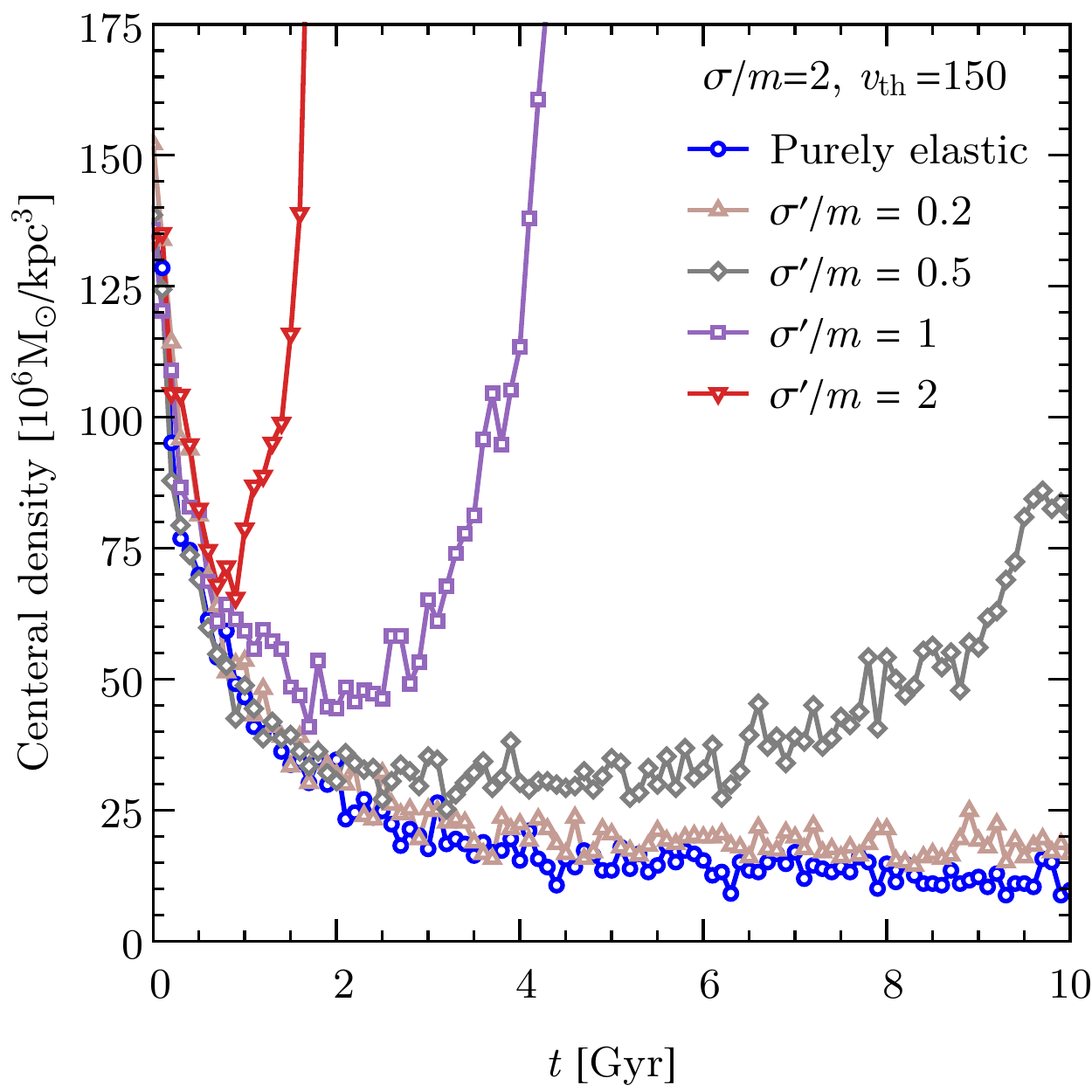}
  \includegraphics[scale=0.46]{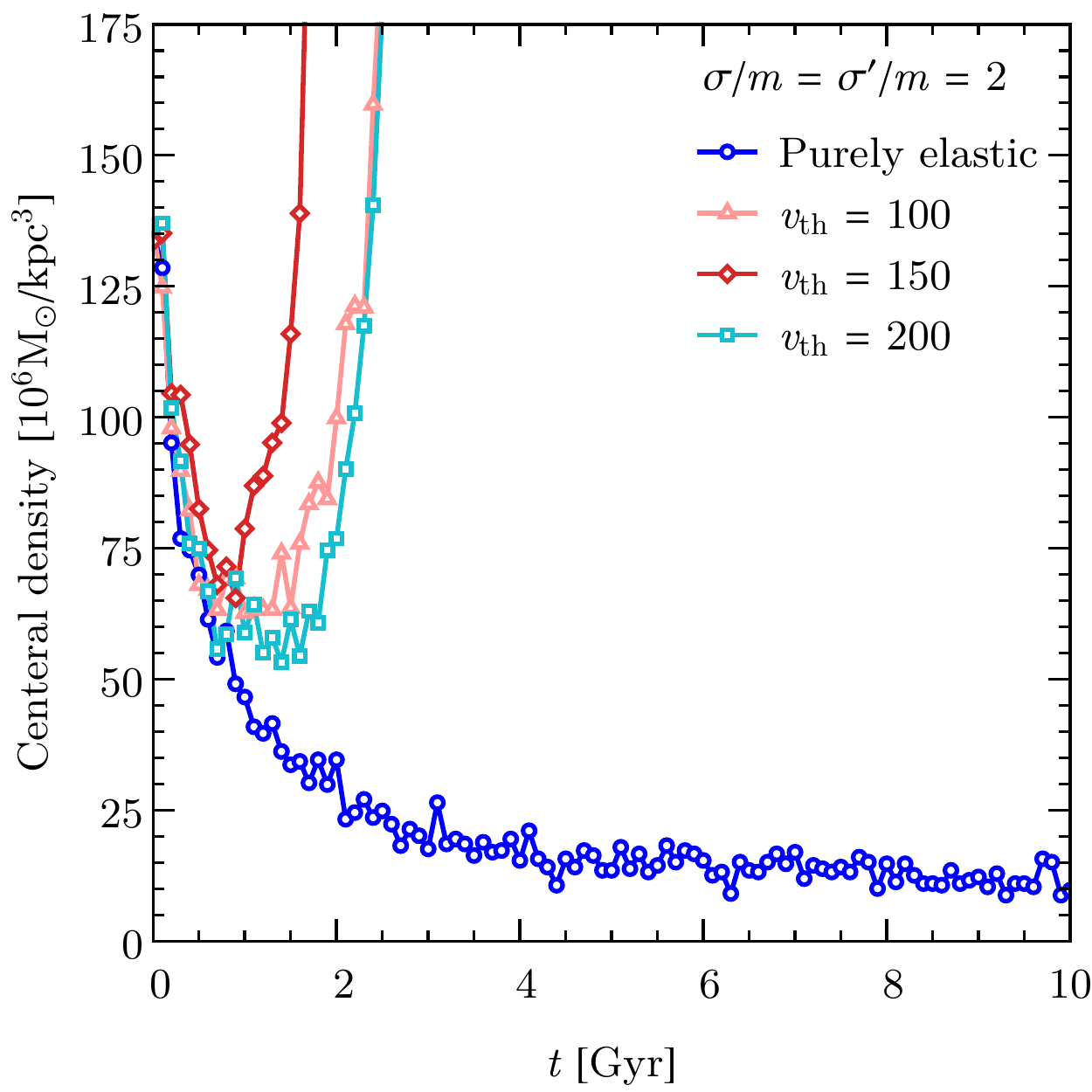}
  \includegraphics[scale=0.46]{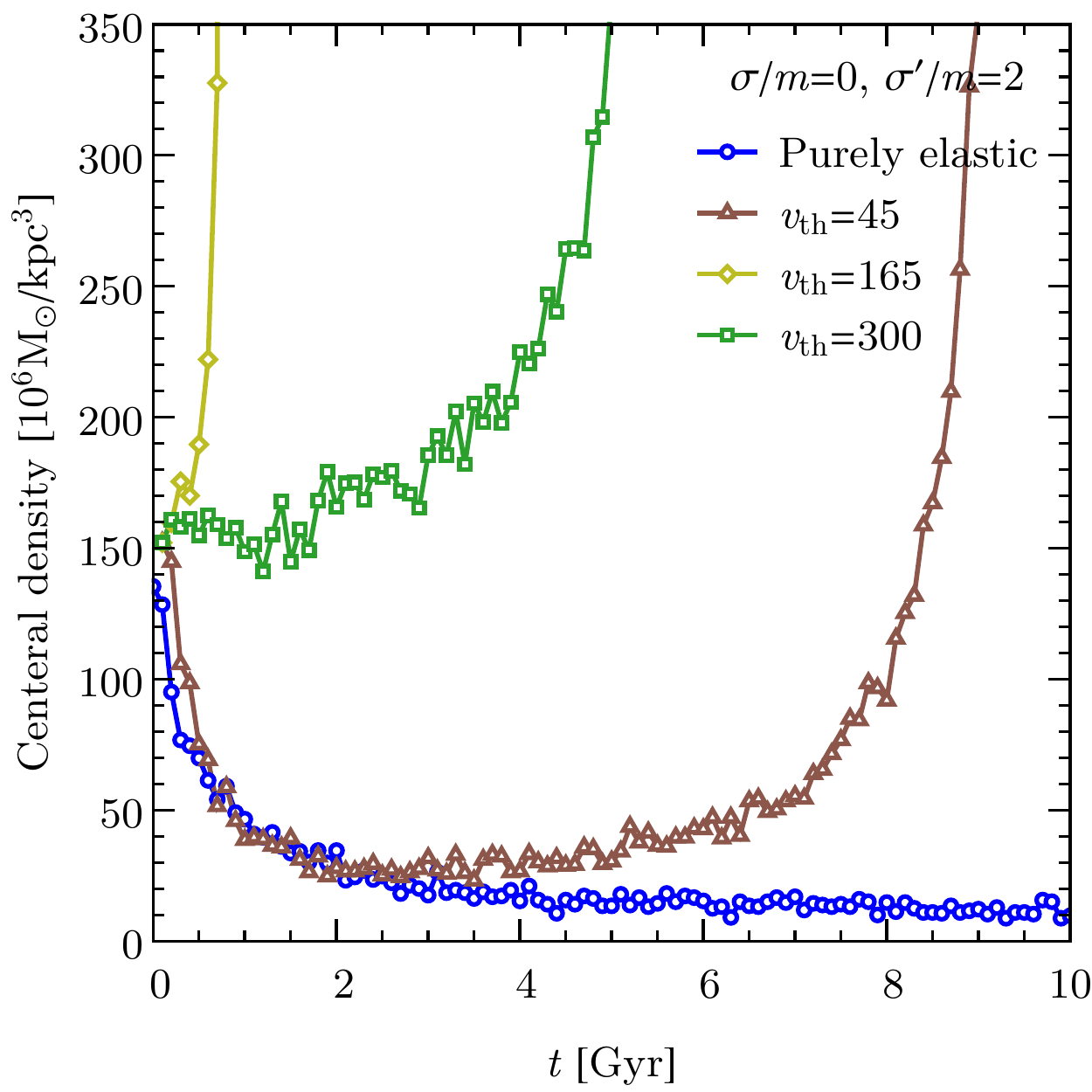}
      \caption{Evolution of the average density within central $1~{\rm kpc}$ for the MW halo with different combinations of model parameters. We vary the dissipative cross section (left), the threshold velocity with (middle) and without (right) the presence of the elastic interactions. The collapse timescale is shortened as the dissipative cross section increases and the threshold velocity becomes comparable to the characteristic velocity of dark matter particles in the halo. In the legends, the cross section per mass and threshold velocity are in units of ${\rm cm^2/g}$ and ${\rm km/s}$, respectively. }
\label{fig:fig2}
\end{figure*}

In our controlled simulations, we choose three benchmark halo masses covering dwarf, MW and cluster scales, and assume Navarro-Frenk-White (NFW)~\cite{Navarro:1995iw,Navarro:1996gj} initial halo profiles. For the dwarf halo, its mass is $M_{\rm vir}=6.28\times10^{10}~{\rm M_\odot}$, the scale radius and density are $r_s = 7.58~{\rm kpc}$ and $\rho_s = 1.17\times 10^7~{\rm M_\odot/kpc^3}$, respectively. For the MW halo, $M_{\rm vir}=1.43\times10^{12}~{\rm M_\odot}$, $r_s = 33.4~{\rm kpc}$ and $\rho_s = 3.13\times10^6~{\rm M_\odot/kpc^3}$. We take $M_{\rm vir}=1.0\times10^{15}~{\rm M_\odot}$, $r_s = 504~{\rm kpc}$ and $\rho_s = 6.25\times10^5~{\rm M_\odot/kpc^3}$ for the cluster halo. We use the public code~\texttt{SpherIC}~\cite{GarrisonKimmel:2013aq} to generate the initial condition with the cutoff radius as $7.8r_s$ for all the halos. The total number of simulated particles is $2\times10^6$ and the gravitational softening lengths are $44~{\rm pc}$, $125~{\rm pc}$ and $1.9~{\rm kpc}$ for dwarf, MW and cluster halos, respectively. We have also performed a convergence test with $4\times10^6$ simulated particles for the MW halo,  and found the difference is minor.

We first study overall features of dissipative SIDM. For all three initial NFW halos, we start with purely elastic simulations with $\sigma/m=2~{\rm cm^2/g}$, then perform dissipative simulations, where we take $\sigma/m=\sigma'/m=2~{\rm cm^2/g}$ and $v_{\rm th}$ values comparable to characteristic dark matter velocities of individual halos, as displayed in Fig.~\ref{fig:fig1}. In addition, to fully explore the impacts of dissipative dark matter self-interactions on halo evolution, we take the MW initial halo as an example and perform a series of simulations with different combinations of $\sigma/m$, $\sigma'/m$ and $v_{\rm th}$. We refer the reader to Fig.~\ref{fig:fig2} for detailed choices of those parameters. 

We also apply the fluid model~\cite{Balberg:2002ue,Essig:2018pzq} to the halos in the purely elastic limit with $\sigma/m=2~{\rm cm^2/g}$, obtain their corresponding density and velocity dispersion profiles, and compare them with our simulation results. Since the fluid model has been carefully calibrated with cosmological SIDM simulations~\cite{Essig:2018pzq}, it provides an independent cross-check of our new SIDM algorithm, as we will show in Fig.~\ref{fig:fig1}.

\section{Density and velocity dispersion profiles}
\label{sec:profiles}

The top panels of Fig.~\ref{fig:fig1} show dark matter density profiles for the dwarf (left), MW (middle) and cluster (right) halos; the bottom panels display their velocity dispersion profiles. We include results from purely elastic and dissipative SIDM simulations, as well as the initial NFW halo profiles. For purely elastic ones with $\sigma/m=2~{\rm cm^2/g}$, dark matter self-interactions transport heat from the outer to inner regions and thermalize the inner halo, resulting in shallow density cores with almost isothermal velocity dispersion profiles for all three halos. We also see our simulations agree with results from the fluid model within $10\%$ in both the density and velocity dispersion. Thus, our SIDM algorithm is further confirmed and validated.

After we turn on the dissipative self-interactions, halo evolution changes dramatically. As shown in Fig.~\ref{fig:fig1} (top), the inner density profiles from dissipative SIDM simulations are much steeper than their corresponding initial NFW profiles. There are two competing effects controlling evolution of the halo. The dissipative interactions tend to reduce the dark matter temperature and lower its velocity dispersion in the inner regions. In turn, this energy loss process can trigger the inner halo to collapse further and convert gravitational energy to kinetic energy, leading to a higher temperature. For the dwarf halo, we take $\sigma/m=\sigma'/m=2~{\rm cm^2/g}$ and $v_{\rm th}=50~{\rm km/s}$ for the dissipative run. Since this $v_{\rm th}$ value is low, dissipative scattering can occur in the whole inner halo. At $t=7~{\rm Gyr}$, the dissipative halo's inner velocity dispersion is lower than the elastic halo's, a clear indication of net energy loss during the evolution. In the purely elastic case, the velocity dispersion has a negative gradient when the core collapse occurs. While in the dissipative case, the dispersion may keep a positive gradient in a collapsed inner halo due to the energy loss~\cite{Essig:2018pzq}. We have also checked that a shallow density core is not observed in the dwarf halo during its entire evolution history.

For the MW halo shown in Fig.~\ref{fig:fig1} (middle), we see that the dissipative self-interactions lead to a steep density profile even at $t=1.9~{\rm Gyr}$ and the associated inner velocity dispersion is higher than the one predicted in the purely elastic case. The temperature enhancement from the core shrink dominates over cooling effects due to energy loss. In Fig.~\ref{fig:fig1} (right), we show simulations for the cluster halo. Similar to the other two halos, the cluster experiences collapse in much less than $10~{\rm Gyr}$ if the dissipative interactions are present. We have checked that MW and cluster halos first enter a short core expansion phase before the collapse occurs. Note for the dissipative simulations shown in Fig.~\ref{fig:fig1}, we have deliberately chosen the threshold velocity to be comparable to the characteristic velocity of dark matter in the halo and maximize the collapse effect, as we will discuss in the next section.

For dark matter models with the dissipative cross section $\sigma'/m\gtrsim 1~{\rm cm^2/g}$, the collapse can occur in much less than $10~{\rm Gyr}$. In the purely elastic limit, taking the initial halo parameters, we use the analytical formula in~\cite{Essig:2018pzq} to estimate the core collapse timescale as ${\cal O}(100)~{\rm Gyr}$. Thus, in dissipative SIDM, the collapse timescale is a factor of $10\textup{--}100$ shorter for the cases we consider. Our $N$-body simulations confirm the prediction based on the semi-analytical fluid model~\cite{Essig:2018pzq}. Furthermore, the collapsed halo induced by the dissipative interactions can have a high central density, but a relatively low temperature due to the energy loss. This is different from the core collapse phenomenon in the purely elastic case, where the temperature always increases towards the center as the collapse occurs.

\section{Halo evolution}
\label{sec:evolution}

Fig.~\ref{fig:fig2} shows evolution of the average density within central $1~{\rm kpc}$ for different sets of MW halo simulations. For the results shown in the left panel, we fix $\sigma/m=2~{\rm cm^2/g}$ and $v_{\rm th}=150~{\rm km/s}$, but vary $\sigma'/m$ in the range of $0.2~{\rm cm^2/g}\textup{--}2~{\rm cm^2/g}$. The threshold velocity is slightly smaller than the halo's average velocity, see Fig.~\ref{fig:fig1} (middle), so most dark matter particles can participate in dissipative collisions. For comparison, we also include the purely elastic simulations, where a large density core quickly forms within $2~{\rm Gyr}$ and it remains for the rest of the evolution. For $\sigma'/m=0.2~{\rm cm^2/g}$, the collisions occur only a few times per particle in the inner halo over the evolution history, and the cooling effect has very minor influence. But it quickly becomes noticeable when $\sigma'/m$ increases. For $\sigma'/m=1~{\rm cm^2/g}$ and $2~{\rm cm^2/g}$, the halo enters the collapse phase around $t=4~{\rm Gyr}$ and $1.5~{\rm Gyr}$, respectively, after a short stage of core expansion. Thus, the collapse timescale is reduced as the dissipative cross section increases.

In the middle panel of Fig.~\ref{fig:fig2}, we show dissipative simulations with $\sigma/m=\sigma'/m=2~{\rm cm^2/g}$ and three different $v_{\rm th}$ values. For all the three cases, the central densities first decrease in the first few ${\rm Gyr}$, but they become cuspy again quickly. Among the three dissipative runs, the one with $v_{\rm th}=150~{\rm km/s}$ has the shortest collapse timescale. As $v_{\rm th}$ increases, the energy loss per up scattering becomes higher, but the number of dark matter particles satisfying $v_{\rm rel}>2v_{\rm th}$ reduces, so the overall cooling rate drops. On the other hand, if $v_{\rm th}$ decreases, more particles can participate in up scattering, but the energy loss per collision becomes smaller. Combing these effects, we can understand that the $v_{\rm th}=100~{\rm km/s}$ and $200~{\rm km/s}$ cases have similar collapse timescales, but both of them are longer than that of the $150~{\rm km/s}$ case.

In the right panel of Fig.~\ref{fig:fig2}, we show {\it purely dissipative} simulations, where we take $\sigma'/m=2~{\rm cm^2/g}$ and vary $v_{\rm th}$. In these simulations, we have turned off elastic dark matter self-scattering completely by setting $\sigma/m=0$; see~\cite{Alvarez:2019nwt} for corresponding particle physics models. For the low threshold velocity, $v_{\rm th}=45~{\rm km/s}$, inelastic scattering can transport the heat and lead to core formation at the first stage of evolution. In this case, although almost all dark matter particles can participate in dissipative collisions, the energy loss per collision is small and the halo can stay in the core expansion phase until around $8~{\rm Gyr}$. It collapses significantly afterwards. While for $v_{\rm th}=165~{\rm km/s}$, the cooling effect is maximized, as $v_{\rm th}$ is close to the halo's characteristic velocity, and the halo enters the collapse phase right after evolution starts and the central density increases monotonically; no shallow density core is observed. 

It is interesting to compare the $v_{\rm th}=165~{\rm km/s}$ case with the dissipative simulations shown in the middle panel, where $\sigma/m=2~{\rm cm^2/g}$ and they all show a short core expansion phase in the beginning. We see that the elastic interactions can help produce a shallow density core and {\it delay} the onset of the collapse triggered by the dissipative interactions. This is because elastic collisions can transport energy to heat up cold patches of the halo and tend to thermalize the inner regions. When we further increase $v_{\rm th}$ to $300~{\rm km/s}$, the energy loss per collision increases, but fewer dark matter particles, which are on high velocity tails, can participate in dissipative up scattering, resulting in a collapse timescale of around $4~{\rm Gyr}$, as shown in the right panel.  

Our simulations show that the collapse timescale is sensitive to the dissipative cross section and threshold velocity; the latter also controls energy loss per dissipative collision. For reasonable choices of the model parameters, the halo can enter the collapse phase in a timescale much less than $10~{\rm Gyr}$. The effect is maximized if the threshold velocity is comparable to the characteristic velocity of dark matter particles in the halo. In some cases, the elastic scattering process redistributes energy and delay core collapse induced by the dissipative interactions.

\section{Conclusion}
\label{sec:conclusion}

We have studied halo structure and evolution in dissipative dark matter. We developed a new SIDM algorithm for $N$-body simulations and verified it in the purely elastic limit. Then, we applied it to model the dissipative interactions and explore their impacts on the halo properties. The interplay between gravitational contraction and collisional dissipation affects the halo evolution in a dramatic way. The dissipative interactions can trigger halo collapse and lead to a steep density profile. The collapse timescale depends on the model parameters that characterize the dissipative interactions in the halo. If the dissipative cross section is around $1~{\rm cm^2/g}$ and energy loss per collision is comparable to halo's typical energy, the collapse can occur within a few ${\rm Gyr}$, much less than the average age of observed galaxies. Overall, our simulations are in good agreement with the results obtained using the modified fluid model in~\cite{Essig:2018pzq}.

Our results have a number of implications. First, it would be important to reevaluate viable parameter space of dark matter models that predict dissipative interactions. If there are multiple states in the dark sector, it is expected to have such interactions. Combining our simulations and observations of dwarf galaxies favoring a density core, we could put novel constraints on those models; see~\cite{Alvarez:2019nwt,Essig:2018pzq}. Second, our simulations do not include baryons. The presence of the baryons can deepen the potential well and further shorten the collapse timescale, as demonstrated in the purely elastic limit~\cite{Sameie:2018chj}. It would be interesting to study predictions of dissipative SIDM models in galaxies with high baryon concentration, such as the MW. Finally, it will be useful to take our simulations and refine the fluid model developed for dissipative dark matter in~\cite{Essig:2018pzq}, which was originally tested against purely elastic SIDM simulations. We will leave these interesting topics for future work.

\acknowledgments
We thank Gerardo Alvarez, Haipeng An and Daneng Yang for useful discussions. This work was supported by the U. S. Department of Energy under Grant No.~DE-SC0008541 (HBY), in part by the Kavli Institute for Cosmological Physics at the University of Chicago through an endowment from the Kavli Foundation and its founder Fred Kavli (YMZ). YMZ thanks the Aspen Center for Physics for hospitality, which was supported by U. S. National Science Foundation grant PHY-1607611. Part of the simulations was run on the Chepfarm computer cluster at Tsinghua University. 

\bibliography{sidm2.bib}

\end{document}